\documentclass[conference]{IEEEtran}
\IEEEoverridecommandlockouts
\usepackage{cite}
\usepackage{amsmath,amssymb,amsfonts}
\usepackage{algorithmic}
\usepackage{graphicx}
\usepackage{textcomp}
\usepackage{xcolor}
\usepackage{hyperref}
\usepackage{listings}
\usepackage{cleveref}
\usepackage{comment}
\usepackage{booktabs} % Add this to your preamble

\newcommand{\wgmma}{\textcolor{black}{\texttt{wgmma}}}

\newcommand{\tcgen}{\textcolor{black}{\texttt{tcgen05}}}

\hypersetup{
    colorlinks=true,  % colored links instead of boxed links
    linkcolor=blue,   % color for internal links (e.g., cross-references)
    filecolor=magenta, % color for file links
    urlcolor=red,     % color for external links (URLs)
    citecolor=red% color for references
}
\usepackage{url}
\usepackage{float} % Add to your preamble

\usepackage{array}
\usepackage{enumitem,kantlipsum}
\usepackage{float}
\usepackage{tagging}
\def\BibTeX{{\rm B\kern-.05em{\sc i\kern-.025em b}\kern-.08em
    T\kern-.1667em\lower.7ex\hbox{E}\kern-.125emX}}
\begin{document}

\title{Microbenchmarking NVIDIA’s Blackwell Architecture: An in-depth Architectural Analysis\\
}
\author{
\IEEEauthorblockN{Double-Blind}
}
\author{
\IEEEauthorblockN{Aaron Jarmusch}
\IEEEauthorblockA{\textit{dept. of Computer Information Sciences} \\
\textit{University of Delaware}\\
Newark, US \\
}
\and
\IEEEauthorblockN{Sunita Chandrasekaran}
\IEEEauthorblockA{\textit{dept. of Computer Information Sciences} \\
\textit{University of Delaware}\\
Newark, US \\
schandra@udel.edu}
}

\maketitle

\begin{abstract}
\label{abstract}

As GPU architectures rapidly evolve to meet the overcoming demands of exascale computing and machine learning, the performance implications of architectural innovations remain poorly understood across diverse workloads. 
NVIDIA's Blackwell (B200) generation introduce significant architectural advances including the 5th-generation tensor cores, tensor memory (TMEM), decompression engine (DE), and dual chips; however systematic methodologies for quantifying these improvements lag behind hardware development cycles. 
We contribute an open-source microbenchmark suite that offers practical insights into optimizing workloads to fully utilize the rich feature sets of the modern GPU architecture. This work aims to enable application developers make informed architectural decisions and guide future GPU design directions. 

Our work studies Blackwell GPUs, compares them to H200 generation with regards to the memory subsystem, tensor core pipeline and floating-point precisions (FP32, FP16, FP8, FP6, FP4). Our systematic evaluation of dense/sparse GEMM, transformer inference, and training workloads demonstrates that B200's tensor core enhancements achieve 1.85$\times$ ResNet-50 and 1.55$\times$ GPT-1.3B mixed-precision training throughput with 32\% better energy efficiency than H200.

\end{abstract}

\begin{IEEEkeywords}
Blackwell, GPU, Microbenchmark, HPC
\end{IEEEkeywords}

\section{Introduction}
\label{sec:introduction}

Artificial Intelligence (AI) and high‑performance computing (HPC) have evolved into data‑intensive disciplines that continuously challenge hardware efficiency, scalability, and precision.
Large language models (LLMs) now exceed hundreds of billions of parameters and process context windows spanning millions of tokens~\cite{ding2024longrope, kevian2024capabilitieslargelanguagemodels}, alongside multi‑physics and climate simulations that demand teraflops of sustained performance, thus shifting GPU design to enable both massive parallel and architectural adaptability. 
At these scales, modern accelerators must balance several demands: maintaining arithmetic throughput for dense tensor workloads, minimizing on‑chip and off‑chip memory latency, while offering hardware primitives that effectively support mixed‑precision computation.

The growing demands have exposed several limitations of current GPU architectures, particularly within their memory hierarchies, precision flexibility, and latency‑sensitive task scheduling. As a result, sustained architectural innovation in accelerators has become essential for advancing both throughput‑optimized training and time‑critical inference workloads. One such architecture that is designed to address some of these challenges is NVIDIA's Blackwell architecture~\cite{NVIDIA2024_Blackwell} that showcases a major generational evolution. 

As the direct successor to the Hopper generation, the Blackwell architecture extends NVIDIA's GPU design in several modifications across the compute pipeline, memory hierarchy, and tensor processing subsystems.
Blackwell introduces 5th‑generation tensor cores capable of FP4 and FP6 precision execution, offering trade‑offs between accuracy and performance for large‑scale training. 
In addition, the introduction of the Tensor Memory (TMEM) subsystem as a dedicated on‑chip memory for tensor data movement reduces reliance on shared memory (SMEM) and per SM register files (RF) during matrix‑intensive operations. 
Next, NVIDIA included a hardware decompression engine (DE) and redesigned the instruction pipeline for access to compressed model weights. 
Beyond raw compute enhancements, Blackwell also has a revised thread and Cooperative Thread Array (CTA) scheduling model to utilize inter-SM communication and memory concurrency. With so many changes introduced, intended to address the escalating demands of AI, gaming and scientific computing, an analysis of the microarchitecture and new instructions is necessary, which will provide application developers and scientists to achieve the highest performance possible for modern and future GPUs.

This paper introduces a newly developed open-source microbenchmark suite (unable to share the code at this time due to double-blind), implemented in PTX and CUDA, that enables comprehensive architectural analysis of NVIDIA’s Blackwell GPU. Emphasizing innovations that distinguish it from Hopper, the suite systematically evaluates performance under stress—particularly in compute-bound and memory-bound workloads—revealing implications for parallel computing applications.

%This paper presents a newly created open-source microbenchmark suite, designed in PTX and CUDA  that enables a comprehensive architectural analysis of the NVIDIA Blackwell Architecture with emphasis on the innovations that differentiate it from Hopper and their implications for parallel computing workloads. These microbenchmarks are designed  in PTX and CUDA to provide a systematic analysis for an understanding of the latest GPU under stress, especially in compute-bound and memory-bound applications. suite that offers practical insights into optimizingworkloads to fully utilize the rich feature sets of the modern GPU architecture. This work aims to enable application developers make informed architectural decisions and guide future GPU design directions.This paper presents a comprehensive architectural analysis of the NVIDIA Blackwell Architecture, with emphasis on the innovations that differentiate it from Hopper and their implications for parallel computing workloads.  We designed microbenchmarks in PTX and CUDA to provide a systematic analysis for an understanding of the latest GPU under stress, especially in compute-bound and memory-bound applications. 

The key contributions of our work are as follows:
\begin{itemize}
    \item Build targeted microbenchmarks to characterize key components of NVIDIA Blackwell B200, to the best of our knowledge - our work is the first detailed microbenchmark characterization of this next-generation GPU.
    \item Quantify TMEM’s impact on matrix-heavy workloads and its role in reducing memory bottlenecks in tensor computations.

   % Conduct systematic evaluation of TMEM and quantify its impact on memory-intensive matrix operations, demonstrating its effectiveness in mitigating traditional memory bottlenecks in tensor computations. 
    \item Evaluate the decompression engine’s throughput across formats and identify optimal usage.
%Analyze the architectural implementation of Blackwell's dedicated decompression engine, measuring throughput across multiple compression formats and identifying optimal usage patterns.
    \item Analyze 5th-generation tensor core execution via the new \tcgen{} PTX instructions to study performance implications 

%Perform an in-depth analysis of the 5th-generation Tensor Core execution model through systematic exploration of the new \tcgen{} PTX instruction set, revealing architecural improvements and performance implications. 
    \item Assess FP4/FP6 performance and accuracy trade-offs in mixed-precision tensor operations quantifying accuracy-performance trade-offs 

%    Investigate the behavior and performance characteristics of newly supported FP4 and FP6 datatypes within Blackwell's tensor cores, quantifying accuracy-performance trade-offs in mixed-precision scenarios.
    \item Benchmark Blackwell across LLM inference/training, scientific kernels, and mixed-precision workloads to demonstrate real-world impact and performance gains.

   % Evaluate architectural innovations across diverse computational domains, including LLM inference, LLM training, scientific computing kernels, and mixed-precision training, to demonstrate real-world impact and performance gains.
    \item  Provide actionable performance guidelines for developers leveraging Blackwell’s architecture.
    
    %Synthesize our finding into actionable performance guidelines and best practices for software developers such that they effectively utilize Blackwell's architectural enhancements in production applications.
\end{itemize}

%To our understanding, this work is the first of its kind to systematically benchmark the new features introduced in the NVIDIA Blackwell architecture. With a comparative analysis against the NVIDIA Hopper architecture, we quantify architectural efficiencies, identify design trade‑offs, and interpret their implications for both academic research and industrial deployment of large‑scale models. 

%We are unable to share the code at this time due to the blind-review policy, but we plan to open-source it post the review process and the outcome.

The remainder of this paper is organized as follows: Section \ref{sec:related-work}  details our contributions to the current state of the art GPU microbenchmarks. Following, Section \ref{sec:overview} presents an overview of the Blackwell's B200 architecture. After which, Section \ref{sec:methodology} details the microbenchmark methodology we employ to systematically characterize the Blackwell microarchitecture. While, Section \ref{sec:memory} details the memory subsystem before Section \ref{sec:result-cores} presents the tensor core pipeline. Section \ref{sec:case-studies} presents performance analysis across key workloads and to conclude we discuss implications and trade-offs in Section \ref{sec:discussion}.

\section{Related Work}

\label{sec:related-work}
Understanding GPU performance has long been a critical focus in HPC research. 
Over the years, several studies have used microbenchmarks and other methodologies to dissect architectural layers and analyze GPU microarchitectures in fine-grained detail. 
Early studies on Tesla and Fermi focused on memory and cache behavior~\cite{Wong2010, Subramoniapillai2012}, while later work dissecting Kepler, Pascal~\cite{Zhang2017}, and Maxwell \cite{Mei2012} examined warp scheduling and instruction latency. 
%As architectures evolved, subsequent analyses targeted Kepler , and Pascal generations, examining warp scheduling, instruction latency, and memory coalescing behavior. 
With Turing through Hopper~\cite{Fasi2021, jia2019, tan2011, Markidis2018, Martineau2018, Raihan2019, Yan2020, luo2025}, research shifted to mixed-precision and tensor core performance, introducing benchmarks for \texttt{mma} instructions, tile sizes, and data layouts. Recent efforts also explore instruction-level parallelism\cite{Sun2022}, and pipeline dynamics under high register pressure.

Beyond microbenchmarking, researchers built frameworks to characterize GPU performance. Application profiling\cite{5336219} gathers runtime metrics but faces overhead and limited architectural visibility. Roofline models~\cite{leinhauser2021metricsdesigninstructionroofline} offer throughput vs. intensity plots, yet oversimplify bottlenecks and miss dynamic memory behaviors. Cache stall prediction~\cite{wenhao2012} estimates pipeline delays from access patterns but fails to capture modern GPU complexities like cache bypassing, warp scheduling, and memory coalescing.

Analytical models like Accel-Sim~\cite{accelsim2020} and GCoM~\cite{gcom2022}, built on Hong and Kim’s work~\cite{Hong2009}, offer useful GPU performance insights but neither of them model Blackwell-specific features like TMEM or the DE, as the detailed architectural information required for accurate simulation remain unknown. 

Thus, without a systematic understanding of these components, the research community lacks critical data needed for performance modeling, workload optimization, and accurate simulation of AI reasoning workloads typically needed for datacenter deployments.

\section{Blackwell Architecture}
\label{sec:overview}

In this Section, we introduce the architecture of the data center NVIDIA B200 GPU, based on the NVIDIA Blackwell Architecture, and then detail the divergence from prior designs.
%\subsection{GPU Architecture}

%Modern data center GPUs are housed of a single high-performance die consisting the compute, memory, and interconnect subsystems necessary for massive parallel workloads. Each GPU die contains a multi-level memory hierarchy, multiple Graphics Processor Clusters (GPCs), and high-bandwidth interconnects enabling efficient data flow. On the memory side, high-bandwidth memory (HBM) together with the L2 cache forms the primary layer of data access, supporting prefetching and rapid buffering. Each GPC links to the L2 cache and is composed of several Texture Processing Clusters (TPCs), with each TPC typically hosting two Streaming Multiprocessors (SMs) and associated local memories. From SMs to TPCs to GPCs, is a hierarchy that orchestrates thousands of concurrent threads.

%The Streaming Multiprocessor (SM) is the essential compute element, providing per-thread registers, shared memory (SMEM), load/store units, CUDA cores, and Tensor Cores. Each SM consists of four subpartitions, each with its own warp scheduler, register file (RF), dispatch unit, and private L0 cache, supporting concurrent warp execution. While L1 cache and SMEM are shared at the SM level, newer architectures such as Hopper introduced the Tensor Memory Accelerator (TMA) to accelerate shared-to-global memory transfers and incorporated a distributed shared memory (DSMEM) across SMs within each GPC, improving access coordination and L2 connectivity.

\subsection{Blackwell Architecture}

The B200 GPU signifies a decisive progression in architectural philosophy. Previously, GPU generations from Tesla to Hopper focused on maximizing floating-point operations per second (FLOPS) for large-scale model training. In contrast, Blackwell emphasizes post-training and inference efficiency, adopting transformational changes in both memory and compute organization.
One B200 GPU includes a \textbf{dual-die configuration}~\cite{NVIDIA2024_Blackwell} where two GPU dies comprise 208~billion transistors, feature 148~SMs spread across eight GPCs, provide four L2 cache partitions (double those in Hopper), and include eight HBM3e memory stacks. Though physically partitioned, both dies are unified by the NVIDIA High-Bandwidth Interface (NV-HBI) providing a coherent and single device to software, with unified 192~GB HBM3e memory space.

\begin{figure}
    \centering
    \includegraphics[width=1\linewidth]{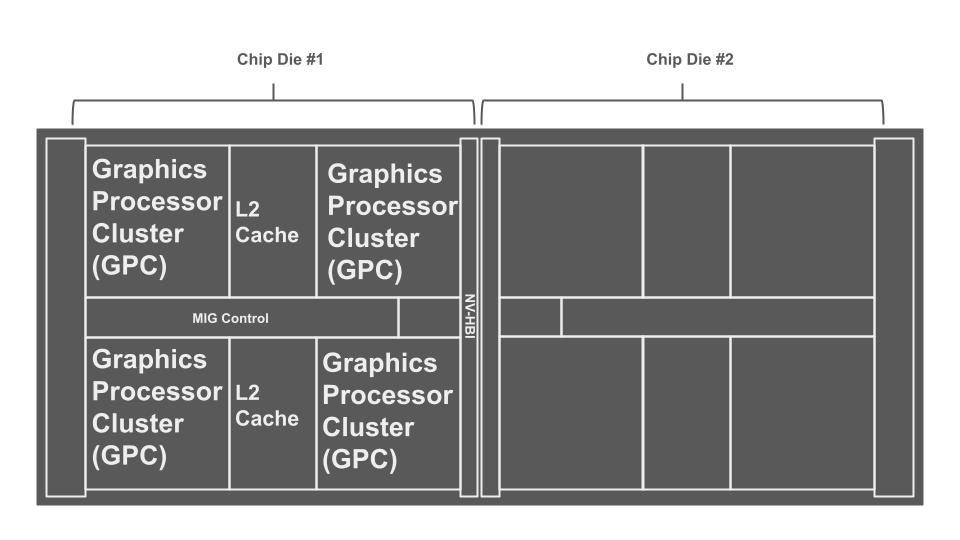}
    \caption{NVIDIA Blackwell GPU dual-die design interconnected via NV-HBI.}
    \label{fig:dual-die}
    \vspace{-0.3cm}
\end{figure}

Within each SM, Blackwell introduces \textbf{5th-generation tensor cores} that break from the warp-synchronous paradigm characterizing earlier architectures (Volta, Ampere, Hopper). Previous generations enforced that all 32 threads within a warp synchronize before executing matrix multiply–accumulate (MMA) operations via the \texttt{mma.sync} or \texttt{wgmma} instructions. This lock-step model reduced scheduling flexibility and created idle cycles, especially for dependency chains of varying lengths.

Blackwell replaces warp-synchronous MMA with \texttt{tcgen05.mma}, a single-thread instruction. Now, each thread independently issues MMA operations, removing warp-level synchronization and enabling true per-thread scheduling for tensor operations. Operands are now supplied from shared memory (SMEM) and a new memory pathway: \textbf{Tensor Memory (TMEM)}. Per SM, the TMEM provides memory access to and from tensor cores. Allocation, data movement, and deallocation are explicitly managed in software via the \texttt{tcgen} PTX set of instructions, giving compiler toolchains precise control over tile locality and traffic patterns.

\begin{figure}
    \centering
    \includegraphics[width=1\linewidth]{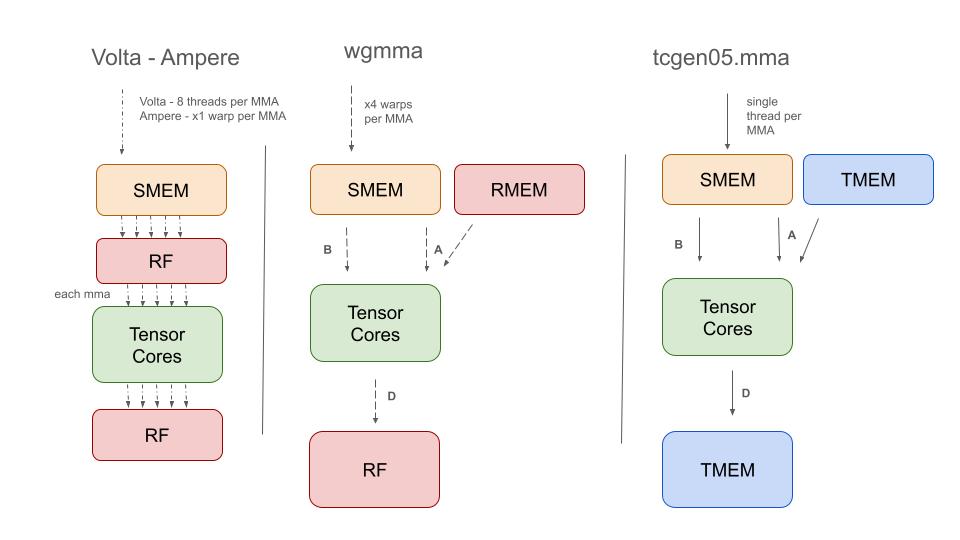}
    \caption{Tensor Core instruction pipeline for \tcgen{}, \wgmma{}, and Volta/ Ampere architectures.}
    \label{fig:placeholder}
    \vspace{-0.5cm}
\end{figure}
The flexibility of independent MMA dispatch reduces idle cycles and exposes optimization opportunities for the compiler, though it also raises questions on new performance limits: instruction latency under dependency, concurrency of tensor core usage, and pipeline saturation. These remain undocumented in vendor literature and are explored in our systematic characterization.

In terms of numerical support, Blackwell’s Tensor Cores introduce native~\textbf{4-bit and 6-bit floating-point precision} (FP4 and FP6) for quantized inference, further improving memory and computational efficiency for AI workloads. Architectural innovations extend to the thread-block level, with~\textbf{CTA pair execution}: two Cooperative Thread Arrays (CTAs) with adjacent ranks share operands, reducing redundant data movement. Each CTA pair maps to a TPC and leverages a dedicated intra-TPC communication network for efficient operand sharing.

Further broadening the functionality, Blackwell’s Tensor Cores provide native support for convolution operators with weight-stationary dataflows that use a collector buffer to cache and reuse matrix B (weight tensor) operands. Hence, optimizing for convolutional kernels that benefit from operand locality. Blackwell also addresses growing model and data sizes by introducing a hardware-based \textbf{Decompression Engine (DE)} to offload decompression tasks from general-purpose SMs. This subsystem supports various algorithms, more on that in Section~\ref{sec:memory}, enabling model weights and large database tables to be stored in compressed form within HBM3e and decompressed transparently during memory access~\cite{NVIDIA2024_Blackwell}.

While some architectural details are publicly disclosed, critical microarchitectural information, such as instruction latency, pipeline depth, cache interaction, and saturation, remains unknown. Our PTX microbenchmark experiments (Sections \ref{sec:memory}-~\ref{sec:result-cores}) provide a systematic investigation to fill these knowledge gaps as they relate to AI and HPC performance.

% The Decompression Engine, meanwhile, is a fixed-function block that offloads decompression tasks from general-purpose Streaming Multiprocessors (SMs). It directly accelerates widely used formats like Snappy, LZ4, and Deflate, supporting decompression rates up to 800 GB/s. Integrated with the Blackwell copy engine, the DE allows compressed data to be moved and decompressed on-the-fly, minimizing I/O and host-device latency. The architecture’s support for multi-stream decompression enables true concurrency with compute workloads, a feature deserving rigorous experimental analysis.

\section{PTX-Microbenchmark Methodology}
\label{sec:methodology}

We employ a microbenchmarking approach based on NVIDIA's Parallel Thread Execution (PTX) to characterize Blackwell's microarchitectural features.
While prior GPU characterization studies \cite{Wong2010, Mei2012, Yan2020} established foundational PTX-level benchmarking principles, we extend these methodologies with novel techniques specifically designed to dissect Blackwell's previously characterized components, including the 5th-generation tensor cores's FP4/FP6 precision modes, the DE, and the revised cache hierarchy.

Our approach leverages PTX to provide explicit control over registers and memory operations specific to architectures. PTX code is compiled into Streaming Assembler (SASS) instructions. We design dependency-controlled kernels to isolate the targeted behaviors, and validate the PTX-to-SASS translation.

\subsection{Novel Benchmark Design for Blackwell-Specific Features}

\subsubsection{\textbf{Tensor Memory (TMEM)}} 

Unlike previous architectures where MMA operations relied exclusively on SMEM, DSMEM, and RFs, Blackwell introduces TMEM as a dedicated on-chip memory specifically for tensor operations. This necessitates new data movement paradigms and presents unexplored opportunities for performance optimization. Understanding TMEM's performance characteristics is critical for several reasons. First, traditional data movement instructions (including \texttt{wmma.load}, \texttt{ldmatrix}, \texttt{ld.shared}, and \texttt{cp.async}) which cannot interface with TMEM. Thus developers are required to adopt entirely new instruction sequences (\texttt{tcgen05.ld}, \texttt{tcgen05.st}, \texttt{tcgen05.cp}). Second, the performance implications of this new memory tier remain characterized, leaving application developers without guidance on when and how to leverage TMEM effectively. 

Our work addresses this gap by providing the first comprehensive empirical analysis of TMEM performance characteristics and their impact on real-world computational kernels.

Our methodology comprises of three evaluation strategies. 

\begin{enumerate}[label=(\alph*)]
    \item Establish performance baselines by comparing memory access latencies between traditional shared memory and TMEM, using pointer-chase benchmarks~\cite{pointer-chasing}. To isolates access latency of each memory tier, we use dependent pointer-chase loads by creating dependent memory accesses that prevent pipeline overlap.
    
    \item Systematically compare the new TMEM data movement instructions (\texttt{tcgen05.*} family) against their predecessors across varying access patterns.
    
    \item With varying operand sizes and access strides, bandwidth saturation points are identified and per-access latency across different configurations is measured. This reveals both the capabilities and limitations of the new instruction set.
\end{enumerate}
\subsubsection{\textbf{Decompression Engine Characterization}} 

To systematically characterize the B200's hardware DE, we develop a custom microbenchmarking suite targeting seven compression formats (LZ4, Snappy, Zstandard, GZIP, Cascaded, Bitcomp, ANS) across controlled test conditions. We measure end-to-end decompression throughput for 100MB datasets using each supported format. Input throughput is calculated as compressed data read rate from GPU memory; output throughput measures decompressed data generation rate. Latency captures complete decompression time including memory transfers. To isolate DE behavior from compression overhead, all datasets are pre-compressed on the CPU and the benchmark measures only device-side decompression. Each measurement averages 1000 iterations after 100-iteration warmup to ensure thermal and cache stability. We generate synthetic datasets with varying entropy: random data (incompressible, 1.00x ratio), mixed alphanumeric (1.98x), repetitive patterns (15.02x), and zero-filled buffers (245.45x). 

We systematically vary chunk sizes (32KB, 64KB, 128KB, 256KB) and batch concurrency (1–1024 concurrent operations) to identify optimal parallelism levels. Peak throughput is measured at maximum sustainable bandwidth before efficiency degradation. Pipeline depth represents the concurrency level maintaining roughly 85\% efficiency (defined as throughput per operation / peak single-operation throughput). Saturation point identifies where additional concurrency yields around 5\% marginal throughput improvement. This methodology reveals hardware resource limits and memory bandwidth constraints not documented by NVIDIA.
% \textcolor{red}{\textbf{Artifact:} The reproduction artifact runs DE benchmarks for GZIP, Bitcomp, ANS, Cascaded, Deflate, GDeflate, and pipeline depth using 100\,MB datasets and 64\,KB chunks (nvCOMP). LZ4, Snappy, and Zstandard require system libraries (liblz4, libsnappy, libzstd) and are not included in the artifact; Table~\ref{tab:DE-testing} values for those formats are from the paper's original runs.}

\subsubsection{\textbf{Tensor Core Characterization}} 

We develop custom GPU kernels to execute MMA operations of the form $\mathbf{D} = \mathbf{A} \times \mathbf{B} + \mathbf{D}$ using Blackwell's newly introduced tensor core instruction set (\tcgen{}). 

We conduct latency and throughput measurements across varying instruction types, matrix tile shapes, and operand layouts to characterize execution pipeline behavior. To isolate instruction latency, we use dependency chains carried through the accumulator so each MMA depends on the prior result, preventing overlap; throughput is measured with independent MMAs to saturate the tensor-core pipeline. Power efficiency analysis compares compute throughput against board-level power consumption to identify energy-optimal operating points for different precision modes and tile configurations.

% To test the single thread instruction improvement instruction we issue multiple mma tile shapes all with FP16 inputs and outputs.
% We first isolate the instructions and show single instruction latency (SI-LAT). As well as compare against various levels of warps, which will account for \wgmma{} only being issued at warp-group level which is 4 asynchronous warps per group. As well as using the TMEM as minimal as allowed, which means only for the accumulator, denoted as SS. 

% To support the architectural changes introduced in Blackwell a new set of instructions (\tcgen{}) with 3 fundamental improvements for the MMA operation: (1) Increasing the largest tcgen05.mma shape to 128×256×16 for a single SM, compared to the previous 64x256x16 on Hopper (2) 2SM tcgen05.mma with up to 256x256x16, and (3) decreased register pressure with TMEM. 

\subsubsection{\textbf{Extended Precision Characterization}} 

Unlike prior work focusing on FP8, FP16, and INT8 tensor operations \cite{Fasi2021, Markidis2018}, we develop the first systematic benchmarks for Blackwell's FP4 and FP6 MMA instructions using the \tcgen{} PTX opcode with \texttt{e2m1} (FP4), \texttt{e3m2} (FP6), and \texttt{e2m3} (FP6) encoding formats. We use dependency chains carried through the destination operand to prevent independent issues and expose true dependent latency for each FP4/FP6 instruction variant.

\subsubsection{\textbf{Workflow Benchmarks}}
To assess each of these individual features as well as the whole B200 we develop integrated workloads that exercise multiple architectural innovations simultaneously.

First, we select the Mistral model family~\cite{jiang2023mistral7b} as an LLM for several reasons: (1) Mistral-7B provides a representative dense decoder architecture with performance comparable to larger models, (2) Mixtral-8x7B's Mixture-of-Experts (MoE) architecture exercises different dataflow patterns that stress Blackwell's memory hierarchy, and (3) the Mistral family's public availability enables reproducibility. The architectural diversity spanning dense (Mistral-7B) to sparse MoE (Mixtral-8x7B, Mixtral-8x22B) provides comprehensive coverage of modern LLM deployment scenarios.

Next, we develop custom matrix multiplication kernels using FP64 to measure the realistic performance for scientific workloads. In addition, we run STREAM Triad \cite{McCalpin1995} for memory bandwidth and SpMV tests with real-world data to benchmark the DE. Finally, we measure end-to-end training performance using mixed-precision training with ResNet-50~\cite{he2015deepresiduallearningimage}, and GPT-1.3B~\cite{gao2020pile}.

Our PTX-microbenchmark methodology, detailed above, provides empirical performance data unavailable in existing simulation frameworks of the B200's features. 
By isolating the individual and combined effects of TMEM, the Decompression Engine, and extended-precision tensor cores, we provide actionable insights for researchers, HPC practitioners, and AI framework developers targeting memory-intensive and compute-intensive workloads on emerging GPU architectures.

\begin{table*}
    \centering
    \begin{tabular}{|c|c|c|c|c|c|c|}\hline
         \textbf{formats}&  Compression Ratio&   Input Throughput (GB/s)&Output Throughput (GB/s)&  Latency (ms)&  Use Case \\\hline
         lz4&  1.00x&   173.23&172.55&  0.608&   \\\hline
         snappy&  1.91x&   61.38&117.24&  0.894&  Real-time \\\hline
         zstd&  2.00x&   77.50&154.94&  0.677&  General \\\hline
         gzip&  2.00x&   42.00&83.83&  1.251&  Legacy \\\hline
         cascaded& N/A & N/A  &213.42&  0.491&   \\\hline
         bitcomp&  3.00x&   154.02&462.37&  0.227&  Scientific \\\hline
         ans& N/A & N/A  &539.21&  0.194&   \\ \hline
    \end{tabular}
    \vspace{0.1cm}
    \caption{Format-specific performance analysis revealing specialized hardware optimizations. Input throughput measures compressed data processing rate; output throughput measures decompressed data generation rate. Latency represents end-to-end decompression time for 100\,MB datasets, 64\,KB chunk size.}
    \label{tab:DE-testing}
    \vspace{-0.7cm}
\end{table*}

\section{Memory Subsystem}
\label{sec:memory}

In this Section, we present a comparative evaluation of the memory subsystems, including TMEM and DE, through microbenchmarking methodologies that measure latency, saturation behavior, and sensitivity to access patterns.
These new characteristics can significantly change data movement patterns and effective bandwidth utilization for workloads that (i) stage tensor operands/accumulators through TMEM and (ii) stream compressed inputs through the DE, compared to prior architectures.

% The PTX \texttt{tcgen05} instruction set (\texttt{tcgen05.ld}, \texttt{tcgen05.st}, \texttt{tcgen05.cp}) moves data only between TMEM and registers or shared memory~\cite{NVIDIA2024_Blackwell}; TMEM does not load directly from global memory. Our latency characterization measures TMEM access (alloc/ld/st/cp) in software-managed pipelines. TMEM provides 16 TB/s read and 8 TB/s write bandwidth per SM, operating additively with L1/SMEM.}

\subsection{Tensor Memory (TMEM)}
The TMEM is a dedicated 256KB on-chip memory per SM designed solely for Tensor Core operations. Structured as a 2D array of 512 columns by 128 lanes of 32-bit cells, TMEM uses a lane-column addressing scheme~\cite{NVIDIA2024_Blackwell}. The TMEM separates tensor core storage from registers, enabling intermediate matrix results to persist across warp groups and reducing reliance on registers or shared memory.

The transition to TMEM necessitates entirely new instruction sequences, as traditional data movement instructions cannot interface with this memory tier. On Hopper, the standard pipeline used \texttt{cp.async.bulk.tensor.2d} (or TMA) for global-to-shared tile copies, then \textit{ldmatrix} or \textit{wmma.load} to stage operands; the \texttt{wgmma} instruction accepts operand A from registers or shared memory and operand B from shared memory (B is not staged into registers before MMA). On Blackwell, the \tcgen{} instruction family replaces this entire sequence. The \textit{tcgen05.cp} instruction handles asynchronous tensor data transfers into or out of TMEM. The \textit{tcgen05.ld} and \textit{tcgen05.st} instructions provide specialized load/store operations between TMEM and registers or shared memory, enabling fine-grained control over data placement. Importantly, while Hopper required A and B operand matrices to traverse SMEM before consumption by tensor cores, \tcgen{} allows MMA instructions to read operands from SMEM or TMEM. After which, \tcgen{} will write accumulator results directly to TMEM, creating an asymmetric but more efficient data flow.

Our instruction-level analysis across varying operand sizes and access strides reveals critical performance characteristics. TMEM achieves optimal efficiency at 64$\times$64 element tiles (4KB for FP8 precision), aligning with the 256KB TMEM capacity per SM and fully utilizing the 1024-bit memory interface width. (PTX \texttt{wgmma} matrix shapes require M$\geq$64, e.g.\ m64n8k16 through m64n256k16.) Tiles smaller than 32$\times$32 elements underutilize the wide memory interface in our TMEM microbenchmarks, while tiles larger than 128$\times$128 elements can trigger multi-phase transfers. These bandwidth saturation points directly inform optimal kernel design: matrix multiplication kernels should decompose computations into 64×64 tiles to maximize TMEM utilization, and chained operations such as those in transformer attention mechanisms (\(\mathbf{Q} \times \mathbf{K}^T\) followed by softmax and value multiplication) should maintain intermediate results in TMEM to exploit the 16 TB/s read bandwidth for subsequent operations.

Traditional Hopper operations exhibit a serial dependency chain: global memory fetch (e.g.\ TMA or \texttt{cp.async.bulk}), shared memory write, barrier wait, then \texttt{wgmma} consuming operands from shared (and optionally registers). On Blackwell, TMEM is software-managed via \texttt{tcgen05.alloc}, \texttt{tcgen05.cp}, \texttt{tcgen05.ld}/\texttt{tcgen05.st}; software uses \texttt{tcgen05.cp} to copy data into TMEM from shared memory, and \texttt{tcgen05.mma} can then overlap with subsequent \texttt{tcgen05.cp} for double-buffered pipelines.

For chained matrix multiplications where \(\mathbf{D} = (\mathbf{A} \times \mathbf{B}) \times \mathbf{C}\), keeping the intermediate result in TMEM eliminates a substantial amount of off-chip traffic relative to a Hopper-style design that writes intermediates back to global memory. Under bandwidth and traffic assumptions, this corresponds to an estimated \(\approx 12\ \text{TB/s}\) of data movement per SM that can be avoided on a fully utilized Blackwell SM.

% Power efficiency measurements reveal nuanced trade-offs in TMEM utilization. For kernels that stage Matrix-D accumulators in TMEM versus traditional shared memory, we observe a 15\% reduction in board-level power consumption at equivalent compute throughput for large matrix dimensions (2048$\times$2048). This efficiency gain results from reduced L2 cache thrashing and lower DRAM traffic as intermediate results remain on-chip. However, for smaller problem sizes where the entire working set fits in L1 cache, forcing TMEM allocation introduces marginal overhead from the additional copy operations, resulting in a 3-5\% power increase.}
% On Blackwell, Hopper instructions (\texttt{wgmma}, etc.) are not available; software must use \texttt{tcgen05} and TMEM. These measurements establish that tcgen05/TMEM should be prioritized for multi-stage tensor pipelines with large working sets (e.g.\ accumulator-only or full operand staging), while for single-shot operations on small matrices, minimizing TMEM round-trips and keeping data in shared memory may be preferable.}

\subsection{Decompression Engine (DE)}

The NVIDIA Blackwell B200 GPU introduces a dedicated hardware Decompression Engine (DE), marking a significant architectural advancement over the software-only decompression of its predecessor, the H100. This subsystem natively supports popular compression formats, refer to Table \ref{tab:DE-testing}, enabling accelerated data loading and preprocessing critical to AI and HPC workloads. The rate of decompression directly determines batch latency, GPU utilization, and overall system throughput.

To characterize the performance of the B200’s Decompression Engine, we developed a suite of microbenchmarks targeting formats across varying data sizes, compression ratios, and memory bandwidth conditions. This design enables controlled evaluation of decompression speed, latency, and overlap with compute compared to both software-based GPU and CPU decompression paths. Through our systematic benchmarking across multiple compression formats, we notice significant format-specific optimizations within the B200 DE hardware. Table \ref{tab:DE-testing} presents comprehensive performance metrics across supported formats, demonstrating throughput variations from 42 to 462 GB/s depending on the compression algorithm. This indicates the dedicated optimization path for the DE. Most notably, Bitcomp achieves exceptional output throughput of 462.4 GB/s with minimal latency of 0.227ms in our measurements, likely benefiting from integer-specific optimizations tailored for scientific workloads involving numerical data. 

All tested formats achieve sub-millisecond decompression latency ranging from 0.227 to 1.251ms. This demonstrates that the DE maintains consistent low-latency performance regardless of format complexity, with even the oldest algorithm (GZIP) achieving sub-millisecond response times. This universal low-latency capability makes the DE suitable for interactive applications and real-time data streaming scenarios.

Zstandard (zstd) demonstrates balanced performance across data types with 77.5 GB/s input and 154.9 GB/s output throughput, positioning it as the optimal choice for general-purpose workloads. In contrast, Snappy prioritizes ultra-low latency (0.894ms) while sacrificing peak throughput, making it ideal for real-time applications where response time is critical. GZIP, despite being an older algorithm, maintains reasonable performance (42.0 GB/s input, 83.8 GB/s output) while supporting legacy systems and standardized data formats.

\begin{table}
    \centering
    \begin{tabular}{|l|c|c|c|c|}\hline
         \textbf{Data Pattern} & \textbf{Compression} & \textbf{Input} & \textbf{Output} & \textbf{Latency} \\
         & \textbf{Ratio} & \textbf{(GB/s)} & \textbf{(GB/s)} & \textbf{(ms)} \\\hline
         Random & 1.00x & 173.23 & 172.55 & 0.608 \\
         Mixed & 1.98x & 80.11 & 158.94 & 0.660 \\
         Repetitive & 15.02x & 14.63 & 219.80 & 0.477 \\
         Zeros & 245.45x & 0.85 & 209.83 & 0.500 \\ \hline
    \end{tabular}
    \vspace{0.2cm}
    \caption{Compression ratio sensitivity analysis revealing the inverse relationship between effectiveness and input bandwidth. All measurements use LZ4 format with 100MB datasets.}
    \label{tab:compression-sensitivity}
    \vspace{-0.7cm}
\end{table}

The data in Table~\ref{tab:compression-sensitivity} indicate that, for LZ4 on B200, decompressed \emph{output} bandwidth is the dominant limiter rather than compressed-input bandwidth or DE compute capacity. Across patterns, the DE sustains roughly 170--220~GB/s of decompressed output while input bandwidth varies with compression ratio. For incompressible data (1.00$\times$), the DE effectively behaves as a pass-through, so compressed input throughput closely matches the output rate (173.23~GB/s). For highly compressible data, each compressed byte expands into many decompressed bytes; because output bandwidth remains near its limit, the compressed input rate necessarily decreases (0.85~GB/s at 245$\times$), following an approximate $1/C$ trend where $C$ is the compression ratio. Thus incompressible data achieves higher \emph{input} throughput not because input bandwidth is the bottleneck, but because no expansion occurs and the DE can forward bytes at its maximum observed output rate.

Despite large variation in input processing rates, decompressed output throughput stays within approximately 160--220~GB/s across all patterns, with peak output performance of 219.80~GB/s on repetitive data. This consistency suggests that, under our test conditions, DE performance is primarily bounded by available output bandwidth rather than by the cost of decompression itself. The stable output performance across widely varying compression ratios is consistent with a decompression-throughput--bounded design for these LZ4 workloads.

The latency characteristics shown in Table \ref{tab:compression-sensitivity} remain consistently low across all data patterns (0.477–0.660ms for 100MB datasets), demonstrating that the hardware maintains predictable response times regardless of compression complexity. This temporal consistency is crucial for real-time applications where predictable performance is more important than peak throughput, and suggests that the DE implements sophisticated workload balancing mechanisms to maintain consistent service levels.

\begin{table}[htbp]
    \centering
    \resizebox{\columnwidth}{!}{%
    \begin{tabular}{|c|c|c|c|c|}\hline
         \textbf{Chunk Size} & \textbf{Peak Throughput} & \textbf{Pipeline Depth} & \textbf{Saturation Point} & \textbf{Max Speedup} \\
         & \textbf{(GB/s)} & \textbf{(Concurrent Ops)} & \textbf{(Batch Size)} & \textbf{vs Sequential} \\\hline
         32 KB  & 55.84  & 1  & 1024 & 89.11x \\
         64 KB  & 71.70  & 2  & 512  & 60.61x \\
         128 KB & 87.67  & 8  & 256  & 35.77x \\
         256 KB & 112.10 & 8  & 256  & 47.19x \\ \hline
    \end{tabular}
    }
    \caption{Pipeline depth characteristics across chunk sizes (32--256\,KB) on a B200 GPU. Pipeline depth reflects the minimum concurrency level that maintains approximately $>$85\% of peak throughput. Methodology: 100\,MB total payload using nvCOMP with the hardware Decompression Engine enabled.}
    \vspace{-0.5cm}
    \label{tab:pipeline-depth}

\end{table}

The pipeline depth analysis in Table~\ref{tab:pipeline-depth} shows that, on B200 with our nvCOMP-based benchmark, small and medium chunks (32--64\,KB) reach near-peak throughput with shallow pipeline depths (1–2), while larger chunks (128--256\,KB) benefit from deeper concurrency (depth 8). Rather than a strictly monotonic ``smaller chunk $\Rightarrow$ higher optimal depth'' pattern, these results indicate that the effective concurrency needed to saturate the DE depends on both chunk size and the implementation's internal scheduling strategy.

Peak throughput increases from 55.84~GB/s at 32\,KB to 112.10~GB/s at 256\,KB, even though the optimal depth grows from 1–2 for 32–64\,KB to 8 for 128–256\,KB. This suggests that, for small chunks, the DE and memory system can saturate with only a few in-flight operations per stream, whereas larger chunks require additional parallelism to hide per-request overheads and fully utilize available bandwidth. Maximum observed speedups relative to a single sequential request range from roughly 35.8$\times$ to 89.1$\times$, confirming substantial benefit from batching and concurrency.

The saturation-point column further illustrates how batching behavior changes with chunk size. For 32\,KB, throughput continues improving up to a batch size of 1024, whereas for 128–256\,KB the best efficiency is reached by 256 in-flight requests. Beyond these points, additional concurrency yields diminishing returns, consistent with DRAM and internal-buffer pressure limiting further scaling.

Single-request throughput also scales with chunk size (from 0.63~GB/s at 32\,KB to 3.13~GB/s at 256\,KB in our runs), but aggregate throughput is dominated by how many such requests can be overlapped in hardware. Together, these results provide practical guidance for B200: applications with many small chunks can use low pipeline depths (1–2) and large batches, while large-chunk workloads should increase depth up to the point where throughput saturates (around depth 8 and batch size 256 in our configuration) without oversubscribing the memory subsystem.

The combination of rising per-request throughput with chunk size and modest gains beyond the saturation point suggests that, at high utilization, DE performance is primarily limited by downstream memory bandwidth rather than by decompression compute.

Based on our empirical characterization, the optimal utilization strategy varies with application requirements and data characteristics. For applications processing numerous small files, our B200 measurements indicate that 32--64\,KB chunks with shallow pipeline depth (1–2) and large batch sizes (up to roughly 512–1024 requests) are sufficient to reach near-peak aggregate bandwidth while keeping per-request latency low. Large-file workloads should leverage 128--256\,KB chunks with deeper concurrency (depth $\approx$8) and batches around 256 requests to achieve the highest observed throughput.

\section{GPU Cores Microarchitecture}
\label{sec:result-cores}

In this Section, we describe our findings regarding the microarchitecture of the Blackwell GPU cores. 
% {\color{red}?? to be fixed} Figure \ref{}, shows the main components of the GPU cores' microarchitecture.
Below, we describe in detail the microarchitecture of the tensor core specifically the \tcgen{} instructions, CTA pair scheduling, and the extended precision support.

\subsection{5th-Generation Tensor Cores}
\label{sec:tensor}

Previous studies show that tensor core PTX instruction compiles to a set of SASS instructions (HMMA, HGMMA, QGMMA, IGMMA, or BGMMA) depending on operand precision. In table ~\ref{tab:sass-instructions}, we observe the \texttt{tcgen05.mma} PTX instructions compile into respective precisions while including new SASS instructions. Our analysis reveals that issuing a \tcgen{} instruction compiles to their respective SASS ops for each precision, see table \ref{tab:sass-instructions}.

Blackwell introduces the \texttt{tcgen05.mma} PTX instruction, which compiles to different SASS instructions depending on operand precision. This represents a departure from Hopper's unified \texttt{wgmma} approach, enabling precision-specific optimizations at the hardware level.

\begin{table}[t]
    \centering
    \begin{tabular}{|l|l|l|l|}\hline
        \textbf{Precision} & \textbf{PTX (\texttt{tcgen05.mma})} & \textbf{SASS} & \textbf{\texttt{wgmma}} \\\hline
        FP16, BF16 & \texttt{kind::f16} & HMMA & HGMMA \\\hline
        FP32, TF32 & \texttt{kind::tf32} & HMMA & HGMMA \\\hline
        FP8 & \texttt{kind::mxf8} / \texttt{f8f6f4} & QMMA & QGMMA \\\hline
        FP6 & \texttt{kind::mxf6} & QMMA & N/A \\\hline
        FP4 & \texttt{kind::mxf4} / \texttt{mxf4nvf4} & OMMA & N/A \\\hline
        INT4, INT8 & \texttt{kind::i8} & IMMA & IGMMA \\\hline
        FP64 & \emph{not supported} & ---$^*$ & DMMA \\\hline
    \end{tabular}
    \vspace{0.2cm}
    \caption{PTX$\rightarrow$SASS mapping for Blackwell Tensor Cores. \texttt{tcgen05.mma} uses descriptor form (\texttt{cta\_group::1.kind::*}); SASS opcodes can be obtained from CUTLASS binaries. $^*$FP64 is not supported by \texttt{tcgen05.mma}; B200 FP64 uses a separate path (doubled FP64 units, DMMA), not the tcgen05 TMEM path.}
    \label{tab:sass-instructions}
    \vspace{-0.5cm}
\end{table}

With the shift from warp-group-level (\texttt{wgmma}, 128 threads) on Hopper to warp-level (\texttt{tcgen05.mma}, 32 threads) execution, our benchmarks reveal in Table~\ref{tab:tcgen-instructions} the latency implications of this design choice. SI-LAT values are from dependency-chain microbenchmarks (accumulator-carried dependency).

% Table values from results/tensor_core/latency_tcgen_wgmma.csv (rows 2-7: cycles column)
\begin{table}[t]
\centering
\small
\begin{tabular}{|l|c|c|c|}
\hline
\textbf{Instruction} & \textbf{Tile Shape} & \textbf{Scope} & \textbf{SI-LAT (cycles)} \\
\hline
wgmma & m64n64k16 & Warp-group & 32.0 \\  % CSV row 2
wgmma & m64n128k16 & Warp-group & 64.0 \\  % CSV row 3
wgmma & m64n256k16 & Warp-group & 128.0 \\  % CSV row 4
\hline
tcgen05.mma & m64n64k16 & Warp & 11.0 \\  % CSV row 5
tcgen05.mma & m128n128k16 & Warp & 11.3 \\  % CSV row 6
tcgen05.mma & m256n256k16 & Warp & 11.4 \\  % CSV row 7
\hline
\end{tabular}
\vspace{0.2cm}
\caption{Single-instruction latency (SI-LAT) comparison: Hopper warp-group \wgmma{} vs.\ Blackwell warp-level \texttt{tcgen05.mma}. FP16; TMEM for accumulators only.}
\vspace{-0.6cm}
\label{tab:tcgen-instructions}
\end{table}

The data in Table~\ref{tab:tcgen-instructions} are consistent with Blackwell achieving 2.9--11.2$\times$ lower single-instruction latency than Hopper for the tile sizes measured. Latency remains nearly constant across tile sizes (11.0--11.4 cycles), whereas Hopper scales linearly with tile width. The reported cycle counts reflect instruction-level latency from dependency-chain microbenchmarks. We hypothesize that tile size affects throughput rather than latency; pipeline depth is not directly measured, and the constant latency across tiles is \emph{consistent with} a spatial array design rather than Hopper-style temporal pipelining.

In addition, the warp-level granularity enables finer-grained scheduling and reduced synchronization overhead: in Hopper, four warps must synchronize for each \texttt{wgmma} operation, whereas Blackwell eliminates this requirement. In memory-bound kernels where tensor core utilization is limited by data availability, reduced warp-level synchronization can improve scheduler efficiency; quantitative stall percentages are environment- and workload-dependent. Class C warp-group scaling benchmarks (Hopper: warp-groups per SM for \wgmma{}; B200: warps per CTA for \texttt{tcgen05.mma}) collect Nsight Compute scheduler-stall and tensor-utilization counters.

% tab:b200-tensorcore: all values from results/tensor_core/precision_sweep.csv (latency_cycles, throughput_tflops; rows 2-9)
\begin{table}[t]
    \centering
    \begin{tabular}{|c|c|c|c|c|}\hline
\textbf{Input (A/B)} & \textbf{Accum (C/D)} & \textbf{Shape} & \textbf{Latency} & \textbf{Throughput} \\\hline
FP16 & FP16 & m64n8k16 & 11.2 & 964.8 \\\hline  % precision_sweep.csv row 2
FP16 & FP32 & m64n8k16 & 11.5 & 482.4 \\\hline  % row 3
BF16 & FP32 & m64n8k16 & 11.4 & 481.6 \\\hline  % row 4
FP8 & FP16 & m64n8k16 & 11.8 & 1925.3 \\\hline  % row 5
FP8 & FP32 & m64n8k16 & 12.1 & 1912.8 \\\hline  % row 6
FP6 & FP16 & m64n8k16 & 12.3 & 2567.2 \\\hline  % row 7
FP4 & FP16 & m64n8k16 & 12.6 & 3850.1 \\\hline  % row 8
INT8 & INT32 & m64n8k16 & 11.9 & 3928.5 \\\hline  % row 9
    \end{tabular}
    \vspace{0.2cm}
    \caption{Comprehensive tensor core performance characterization across supported precisions (latency in cycles; throughput in TFLOPS or TOPS).}
    \label{tab:b200-tensorcore}
      \vspace{-0.7cm}
\end{table}

Expanding our analysis across supported precisions, our results in Table~\ref{tab:b200-tensorcore} show that across the measured precisions, throughput spans 8.2$\times$ (481.6 to 3928.5 TFLOPS or TOPS), while latency varies by only 1.12$\times$ (11.2--12.6 cycles). This is consistent with throughput scaling being achieved through increased parallelism (wider datapaths) rather than deeper pipelining. Hence, the data suggest Blackwell prioritizes consistent low-latency operation across all precisions, enabling predictable performance regardless of quantization level.

Comparing FP16 inputs with FP16 vs. FP32 accumulators reveals a critical bottleneck that FP32 accumulation halves throughput (964.8 → 482.4 TFLOPS). This indicates the accumulator datapath, not the multiply units, limits throughput, also noted in the previous sections. Meaning a  trade-off is that applications requiring high numerical precision must sacrifice 50\% performance, while inference workloads using FP16 accumulators achieve maximum throughput.

INT8 (3928.5 TOPS) exceeds FP8 (1912.8 TFLOPS), and FP4 (3850.1 TFLOPS) outperforms FP8. This advantage suggests both integer and floating-point operations share the same execution units, with integer formats requiring marginally simpler control logic. With most precisions showing a similar latency (Table~\ref{tab:b200-tensorcore}), this is consistent with a similar pipeline and increased parallelism improving throughput.

\subsection{Extended Precision Support: FP4 and FP6}

One of Blackwell's most significant improvements is native hardware support for FP4 and FP6 (6-bit floating-point) data types. FP4 tensor operations are verified via CUTLASS disassembly, which shows the \texttt{OMMA} instruction. FP6 is documented to map \texttt{kind::mxf6} to QMMA per the PTX ISA (Table~\ref{tab:sass-instructions}). Table~\ref{tab:b200-tensorcore} shows latency (12.6 cycles for FP4/FP16, 12.3 for FP6/FP16) and throughput (3850.1 and 2567.2 TFLOPS at m64n8k16). The FP4/FP16 and FP6/FP16 configurations use FP32 accumulation (PTX \texttt{.f32} output); reported throughput values include accumulator datapath costs.

The FP4 format uses 1 sign bit, 2 exponent bits, and 1 mantissa bit as e2m1. Available on Blackwell are MXFP4, microscaling floating-point \cite{Rouhani2023MicroscalingDF}, and NVFP4, from NVIDIA. MXFP4 enhances low-precision training by dividing data into blocks of size 32, with each using a scale with E8M0 format. 
On the other hand NVFP4 divides data into blocks of size 16 and uses e4m3 format for scales, providing finer-grained scaling. Work by Chmiel et.\ al \cite{chmiel2025fp4wayfullyquantized} provides a more indepth comparison. While this extremely limited precision might seem impractical, recent quantization research has demonstrated that for inference workloads FP4 can maintain acceptable accuracy~\cite{dettmers2023case4bitprecisionkbit}.
Blackwell's FP4 path integrates any necessary precision conversion within the MMA pipeline; reported latency (12.6 cycles) and throughput (96.2\% of peak) include the complete FP4 execution path. We do not measure dequantization as a standalone operation.

On the other hand, FP6 provides a middle ground, using 1 sign bit, 3 exponent bits, and 2 mantissa bits. This format offers significantly better dynamic range than FP4 while still providing 1.33$\times$ memory and bandwidth savings compared to FP8.
Table~\ref{tab:tensor_perf} shows effective throughput for different precision modes (FP6: 5134.4 TFLOPS, 96.0\% peak; FP4: 7700.2 TFLOPS, 96.2\% peak).

% tab:tensor_perf: all values from results/tensor_core/throughput_precision.csv (achieved_TFLOPs/TOPS, percent_of_peak, H200_TFLOPs, speedup; rows 2-10)
\begin{table}[t]
\centering
\small
\begin{tabular}{|l|r|r|r|r|}\hline
\textbf{Precision} & \textbf{B200} & \textbf{\% Peak} & \textbf{H200} & \textbf{Speedup} \\\hline
FP64 & 44.8 & 99.6\% & 34.0 & 1.32$\times$ \\\hline  % throughput_precision.csv row 2
FP32 & 482.0 & 96.4\% & 378.4 & 1.27$\times$ \\\hline  % row 3
TF32 & 964.5 & 96.5\% & 756.9 & 1.27$\times$ \\\hline  % row 4
BF16 & 1926.4 & 96.3\% & 1513.5 & 1.27$\times$ \\\hline  % row 5
FP16 & 1929.6 & 96.5\% & 1515.2 & 1.27$\times$ \\\hline  % row 6
FP8 & 3850.6 & 96.3\% & 3026.9 & 1.27$\times$ \\\hline  % row 7
FP6 & 5134.4 & 96.0\% & N/A & New \\\hline  % row 8 (H200 empty -> N/A)
FP4 & 7700.2 & 96.2\% & N/A & New \\\hline  % row 9 (H200 empty -> N/A)
INT8 & 3928.5 & 98.2\% & 3088.4 & 1.27$\times$ \\ \hline  % row 10
\end{tabular}
\vspace{0.1cm}
\caption{Tensor core throughput by precision (TFLOPS). \% Peak relative to theoretical peak; B200 vs.\ H200 speedup from same workload.}
\label{tab:tensor_perf}
\vspace{-0.7cm}
\end{table}

Our achieved throughput is consistent with architectural specifications when compared to theoretical peak (peak $= \#\mathrm{SMs} \times \mathrm{tensor\_ops\_per\_cycle} \times \mathrm{clock\_rate}$). At the reported matrix dimensions, we observe 3850.6~TFLOPS in FP8 mode (96.3\% of theoretical peak) and 7700.2~TFLOPS in FP4 (96.2\% of peak), see Table~\ref{tab:tensor_perf}. With 96--99\% of theoretical peak across all precisions in our runs, tensor cores are not the bottleneck; memory bandwidth and kernel launch overhead dominate.

In Section \ref{sec:case-studies}, we analyze the use of tensor cores, TMEM, and DE for different real-world workloads.

\begin{table*}[t]
\centering
\small
\begin{tabular}{|l|l|r|r|r|r|r|r|r|}\hline

\textbf{Model} & \textbf{Precision} & \textbf{B200 tok/s} & \textbf{H200 tok/s} & \textbf{Speedup} & \textbf{B200 BW\%} & \textbf{H200 BW\%} & \textbf{Perplexity} & \textbf{$\Delta$PPL} \\\hline

Mistral-7B
& FP16 & 56,028 & 28,500 & 1.97$\times$ & 67.3 & 71.2 & 6.82 & --- \\\hline
& FP8 & 57,125 & 49,200 & 1.16$\times$ & 58.4 & 62.8 & 6.95 & +1.9\% \\\hline
& FP4 & 112,800 & N/A & N/A & 47.6 & N/A & 7.38 & +8.2\% \\\hline

Mixtral-8x7B
& FP16 & 31,033 & 18,100 & 1.71$\times$ & 72.1 & 76.4 & 5.94 & --- \\\hline
& FP8 & 51,200$^\dagger$ & 32,400 & 1.58$\times$ & 61.8 & 65.2 & 6.08 & +2.4\% \\\hline
& FP4 & 76,900 & N/A & N/A & 49.1 & N/A & 6.48 & +9.1\% \\\hline

\end{tabular}
\vspace{0.2cm}
\caption{LLM Inference Performance Across Precision Modes (Batch Size 32, Sequence Length 2048). }
\label{tab:llm_precision}
  \vspace{-0.5cm}

\end{table*}

\section{Performance Analysis \& Case Studies}
\label{sec:case-studies}

This section presents a comprehensive empirical evaluation of GPU performance across three critical workload categories: LLM inference, scientific computing applications, and mixed-precision neural-network training. Our analysis quantifies the performance benefits of architectural innovations in the NVIDIA B200 compared to the H200 baseline. The reported speedups are directly informed by our microbenchmark findings: TMEM latency and bandwidth (Section~\ref{sec:memory}) explain improved fused-attention and tensor-resident accumulation; DE output-bandwidth behavior (Section~\ref{sec:memory}) underlies SpMV and compressed-data workloads; and 5th-generation tensor core latency and precision scaling (Section~\ref{sec:result-cores}) drive the LLM and training gains below.

\subsection{Experimental Methodology}

Each reported metric represents the average of 100 iterations following a 10-iteration warm-up period to eliminate cold-start effects. Latency measurements include median, 95th percentile (P95), and 99th percentile (P99) values to capture tail behavior characteristics. Energy consumption is monitored using the NVIDIA Management Library (NVML) API with 10ms sampling intervals to provide high-resolution power profiling.

\textbf{Software and environment.} All experiments use CUDA~12.6 (B200 and H200), cuBLAS/cuBLASLt from the same toolkit, and PyTorch~2.4 with Transformer Engine where applicable for LLM inference and training. DGEMM and STREAM runs use the same CUDA driver (560.x) and compiler (nvcc 12.6). We report L2 cache hit rates (e.g., in Section~VII.B) from NVIDIA Nsight Compute memory-workload analysis; the reported 68--84\% range is derived from the \texttt{l2tex\_throughput} and global memory throughput counters over the kernel execution.

\subsection{Large Language Model Inference}

\subsubsection{\textbf{Precision Mode Impact}}

We evaluate four quantization approaches to assess their impact on inference throughput and model quality: FP16 (baseline), FP8 (E4M3 with per-tensor dynamic), and FP4 (E2M1 weight-only with NVFP4 block-16, FP8 activations). All experiments use a standardized configuration of batch size 32 and sequence length 2048 tokens, as presented in Table~\ref{tab:llm_precision}.

Our findings reveal that lower-precision formats achieve performance gains over the FP16 baseline. Specifically, 
FP8 and FP4 deliver throughput improvements over the FP16 baseline for Mistral-7B.
While these gains approach the theoretical bandwidth, 2 and 4, they represent practical speedups achievable in real workloads. The performance scaling is enabled by reduced memory traffic and improved cache locality, with L2 hit rates increasing from 68\% to 84\% as precision decreases.
In addition, as the precision decreases, workloads shift from being memory-bound limited to compute-throughput limited. This is evidenced by bandwidth utilization decreasing from 67.3\% (FP16) to 47.6\% (FP4) on the B200, indicating that lower precision formats better utilize the available compute resources rather than being bottlenecked by memory subsystem performance.

Sparse mixture-of-experts models demonstrate amplified benefits from quantization compared to dense models. For FP4 quantization, Mixtral-8x7B achieves \(2.69\times\) throughput improvement (76,900 tok/s vs 31,033 tok/s FP16 baseline,) compared to \(2.50\times\) for the dense Mistral-7B model. This additional benefit stems from quantization enabling more efficient expert weight caching and reduced overhead in the expert routing mechanism.

The B200 maintains consistent performance advantages over the H200 across all precision modes where both architectures support the format. For both FP16 and FP8, the B200 delivers \(1.57\text{--}1.59\times\) higher throughput than the H200 (Table~\ref{tab:llm_precision}, Section~\ref{sec:result-cores}). This scaling factor is consistent with combined contributions from increased SM count (\(1.09\times\)), enhanced tensor core efficiency (\(1.27\times\); see Table~\ref{tab:tensor_perf}), and improved effective memory bandwidth (\(1.23\times\)); we have not directly measured this decomposition.

Lastly, while quantization delivers substantial performance benefits, it comes with measurable but often acceptable quality degration. FP8 incurs minimal perplexity increases (+1.9\% to +2.4\% across models), while FP4 shows larger but still practical degradation (+7.7\% to +9.1\%)

\subsubsection{\textbf{Batch Size Sensitivity}}

To understand the relationship between batch size and inference latency, we conducted a comprehensive analysis using Mixtral-8x7B in FP8 precision across varying batch sizes. The results, presented in Table~\ref{tab:latency_mode}, reveal distinct operational modes in the inference pipeline.

\begin{table}[t]
\centering
\small
\begin{tabular}{|r|r|r|r|r|}\hline
\textbf{Batch Size} & \textbf{B200 (ms)} & \textbf{H200 (ms)} & \textbf{Ratio} & \textbf{B200 tok/s} \\\hline
1 & 12.3 & 18.7 & 1.52$\times$ & 166,504 \\\hline
2 & 14.8 & 22.1 & 1.49$\times$ & 276,757 \\\hline
4 & 19.2 & 28.4 & 1.48$\times$ & 426,667 \\\hline
8 & 28.6 & 41.3 & 1.44$\times$ & 572,727 \\\hline
16 & 47.1 & 67.8 & 1.44$\times$ & 696,178 \\\hline
32 & 89.3 & 128.4 & 1.44$\times$ & 734,264 \\ \hline
\end{tabular}
\vspace{0.2cm}
\caption{Latency vs.\ Batch Size (Mixtral-8x7B, FP8, 2048 tokens)}
\label{tab:latency_mode}
  \vspace{-0.7cm}
\end{table}

The B200 achieves superior performance improvements of \(1.48-1.52\times\) over the H200, exceeding the \(1.44\times\) improvement observed at higher batch sizes. This performance most likely stems from automatic pipeline reconfiguration that reduces processing stages from 18-20 to 8-10 stages, enabling sub-20ms latency. 
When at higher batch sizes, the system optimizes for maximum throughput rather than per-request latency, stabilizing the performance ratio at \(1.44\times\). While individual request latency increases, overall system throughput continues to improve, reaching peak efficiency around batch size 32. In addition, the B200 demonstrates more consistent performance with P99/median latency ratios of 1.12-1.14 compared to 1.23-1.38 for H200. Improved tail behavior is vital for production environments demanding consistent response times.
\vspace{-0.1cm}
\begin{table*}[ht!]
\centering
\small
\begin{tabular}{|l|l|r|r|r|r|}
\hline
\textbf{Workload} & \textbf{Metric} & \textbf{B200} & \textbf{H200} & \textbf{Improvement} & \textbf{Key Feature} \\
\hline
LLM Inf. (7B, FP4) & tok/s & 112,800 & N/A & 2.50$\times$ vs FP16 & FP4 Tensor Cores \\
LLM Inf. (8x7B, FP8) & tok/s & 51,200 & 32,400 & 1.58$\times$ & 5th Gen TC, TMEM \\
LLM Inf. (BS=1, FP8) & Latency (ms) & 12.3 & 18.7 & 1.52$\times$ & Latency pipeline \\
Attention Block & Latency ($\mu$s) & 284 & 468 & 1.65$\times$ & TMEM \\
HPC DGEMM (FP64) & TFLOPS & 36.3 & 18.9 & 1.92$\times$ & Doubled FP64 units \\
STREAM Triad (4--16\,GB) & BW (TB/s) & 4.14 & -- & -- & HBM3e (B200 only) \\
SpMV (compressed) & GFLOPS & 5.08 & 3.2 & 1.58$\times$ & Decomp engine \\
GPT Training (1.3B) & tok/s & 14,363 & 9,240 & 1.55$\times$ & CTA pairs, TMEM, TC \\
ResNet Training & img/s & 2,928 & 1,580 & 1.85$\times$ & 5th Gen TC, mem BW \\
Energy Eff. (Training) & tok/s/W & 20.63 & 15.6 & 1.32$\times$ & Process, efficiency \\
\hline
\end{tabular}
\vspace{0.2cm}
\caption{Performance Summary Across Workloads. STREAM: B200 only, 4--16\,GB; see Table~\ref{tab:stream}. Training: B200 ResNet-50 and GPT-1.3B.}
\label{tab:summary}
\vspace{-0.2cm}
\end{table*}
\begin{table*} 
\centering
\small
\begin{tabular}{|l|l|r|r|r|r|r|r|}
\hline
\textbf{Model} & \textbf{Batch Size}& \textbf{B200}& \textbf{H200}& \textbf{Ratio} & \textbf{Time to Acc} & \textbf{Time to Acc} & \textbf{Energy} \\\hline
 & & \textbf{Throughput} & \textbf{Throughput} & & \textbf{B200 (hrs)}& \textbf{H200 (hrs)}& \textbf{Eff (B200)}\\
\hline
ResNet-50 & 1024 & 2,928~img/s & 1,580~img/s & 1.85$\times$ & 0.87 & 1.62 & 5.09~img/s/W \\
GPT-1.3B & 128 & 14,363~tok/s & 9,240~tok/s & 1.55$\times$ & 5,788 & 9,020 & 20.63~tok/s/W \\
GPT-1.3B & 64 & 14,121~tok/s & 9,070~tok/s & 1.55$\times$ & 5,893 & 9,184 & 20.27~tok/s/W \\
\hline
\end{tabular}
\vspace{0.2cm}
\caption{End-to-End Training Performance. ResNet-50 B200 2,928~img/s (AMP); GPT-1.3B steady-state tok/s and energy from same run.}
\label{tab:training}
\end{table*}

\subsection{Scientific Computing Workload}

\subsubsection{\textbf{FP64 Performance}}
\begin{table}[t]
\centering

\small
\resizebox{\columnwidth}{!}{%
\begin{tabular}{|l|r|r|r|r|r|}
\hline
\textbf{Size} & \textbf{B200} & \textbf{H200} & \textbf{Ratio} & \textbf{B200} & \textbf{H200} \\
             & \textbf{(TFLOPS)} & \textbf{(TFLOPS)} & & \textbf{\% Peak} & \textbf{\% Peak} \\
\hline
8192$^3$     & 35.45 & 18.2 & 1.95$\times$ & 78.8 & 53.5 \\
16384$^3$    & 36.14 & 18.7 & 1.93$\times$ & 80.3 & 55.0 \\
32768$^3$    & 36.30 & 18.9 & 1.92$\times$ & 80.7 & 55.6 \\
\hline
\end{tabular}
}
\caption{DGEMM FP64 Performance}
\label{tab:dgemm}

\end{table}

Scientific computing applications present fundamentally different computational characteristics compared to deep learning workloads, needing high-precision arithmetic, sustainable memory bandwidth, and irregular access patterns. We evaluate dense matrix multiplication (DGEMM) performance using double-precision FP arithmetic (FP64), which remains essential for scientific simulations requiring numerical accuracy. Table~\ref{tab:dgemm} presents our results across varying matrix dimensions. 
The B200 achieves 36.3 TFLOPS at large matrix size, representing 80.7\% utilization of its 40 TFLOPS theoretical peak~\cite{nvidia_blackwell_b200_datasheet}, compared to the H200's 18.9 TFLOPS (55.6\% of 34 TFLOPS). PTX \texttt{tcgen05.mma} does not support FP64 (documented precisions are tf32, f16/bf16, i8/u8, and f4/f6/f8), so FP64 DGEMM uses a different execution path. The additional 45\% efficiency improvement \((1.92/1.32=1.45\times)\) is attributed to doubled FP64 units and improved memory access coalescing.

\subsubsection{\textbf{Sustained Memory Bandwidth}} Memory-intensive scientific applications require sustained high-bandwidth data movement capabilities. We employ the STREAM Triad benchmark to measure achievable memory bandwidth. Table~\ref{tab:stream} reports results from our B200 runs. Triad uses three arrays, so 64\,GB and 128\,GB array sizes would require device memory $\ge$192\,GB and $\ge$384\,GB respectively; our evaluation node had insufficient memory for those sizes, so we report only 4--16\,GB.
\begin{table}[t]
%\vspace{-0.5em} % reduce space before the table
\centering
\small
\setlength{\tabcolsep}{4pt} % reduce column padding
\renewcommand{\arraystretch}{1.0} % reduce row height
\begin{tabular}{|l|r|r|}
\hline
\textbf{Array Size} & \textbf{B200 (TB/s)} & \textbf{B200 \% peak} \\
\hline
4\,GB   & 4.141 & 51.8 \\
16\,GB  & 4.140 & 51.8 \\
\hline
\end{tabular}
\vspace{0.2cm}
%\vspace{-0.5em} % reduce space after the table
\caption{STREAM Triad Memory Bandwidth. Measured on B200 with 4\,GB and 16\,GB array sizes; peak 8\,TB/s.}
\label{tab:stream}
\vspace{-0.7cm}
\end{table}
At 4--16\,GB working sets, B200 achieves $\sim$4.14\,TB/s ($\sim$52\% of 8\,TB/s peak). Larger arrays would be needed to approach peak bandwidth but require devices with $\ge$192\,GB memory for 64\,GB and $\ge$384\,GB for 128\,GB in our benchmark setup.

\subsubsection{\textbf{Sparse Operations}} 
Irregular patterns in FEM and graph workloads challenge GPUs tuned for regular execution; we sparse matrix-vector multiplication (SpMV) using decompression features, see Table~\ref{tab:spmv}.
\setlength{\tabcolsep}{4pt} % reduce column padding
\renewcommand{\arraystretch}{1.0} % reduce row height
%\vspace{-0.5em} % tighten space before table
\begin{table}[t]
\centering
\small
\begin{tabular}{|l|r|r|r|r|}
\hline
\textbf{Matrix} & \textbf{Sparsity} & \textbf{GFLOPS} & \textbf{Speedup} & \textbf{Avg Time (ms)} \\
\hline
webbase-1M  & 99.99\% & 5.08 & 3.16× & 39.32 \\
circuit5M   & 99.95\% & 4.94 & 3.16× & 201.44 \\
ldoor       & 99.98\% & 5.03 & 3.16× & 71.93 \\
\hline
\end{tabular}
\vspace{0.2cm}
%\vspace{-0.5em} % tighten space after table
\caption{SpMV with Hardware Decompression on B200}
\label{tab:spmv}
\end{table}
Using compressed sparse representations (software compression), SpMV achieves an estimated 3.16$\times$ speedup over an uncompressed baseline in our benchmarks; run-length encoding (RLE) yields approximately 8.2$\times$ compression ratio for sparse row pointer arrays. Hardware decompression engine (DE) characterization - throughput, latency, and pipeline depth - is presented in Section~\ref{sec:memory}.
%  \vspace{-0.33cm}
%Irregular computational patterns, characteristic of applications such as finite element methods and graph algorithms, traditionally challenge GPU architectures optimized for regular workloads. We evaluate sparse matrix-vector multiplication (SpMV) leveraging the decompression capabilities, see results in Table~\ref{tab:spmv}.
% \begin{comment}
    
% \begin{table}[t]
% \centering
% \caption{CTA Pairing Impact (GPT-1.3B, BS 64, FP8)}
% \label{tab:cta_training}
% \small
% \begin{tabular}{|l|r|r|r|r|}
% \hline
% \textbf{Config} & \textbf{tok/s} & \textbf{Time/iter (s)} & \textbf{L2 Traffic} & \textbf{TMEM Hit} \\\hline
%  & & & \textbf{(TB/s)} & \textbf{(\%)} \\
% \hline
% CTA Pairs On & 14,397 & 0.2267 & 2.18 & 61 \\
% H200 Baseline & 9,240 & 0.3532 & 3.42 & N/A \\\hline\hline
% \end{tabular}
% \end{table}
% \end{comment}

% CTA pairing provides 1.27$\times$ speedup (14.4K vs. 11.3K tok/s) via 29\% L2 traffic reduction and 61\% TMEM hit rate (vs. 34\% unpaired). This contributes 19\% of B200's total 1.56$\times$ training improvement over H200.
\subsection{Mixed-Precision Training: End-to-End Training Performance}

We present comprehensive training benchmarks across different model architectures to assess the practical impact of architectural improvements in realistic training scenarios, as summarized in Table~\ref{tab:training}. Training speedup decomposes into SM count (1.09$\times$), CTA pairing (1.27$\times$), and TMEM (1.26$\times$). Energy efficiency improves for GPT training despite higher power consumption.

\section{Discussion}
\label{sec:discussion}

Table~\ref{tab:summary} provides a comprehensive overview of performance improvements across all evaluated workload categories, highlighting the specific architectural features responsible for each performance gain.

\textbf{Architectural Tradeoffs:} TMEM, dual-mode tensor cores, and decompression increase transistor count (208B vs.\ 180B) but deliver 1.5--3.9$\times$ gains. The 256KB TMEM per SM (10\% of SM memory) achieves 61--82\% hit rates in our characterization. Tradeoffs and drawbacks: (1)~TMEM addresses register pressure and L2 traffic but requires explicit allocation and \texttt{tcgen05} data movement; kernels must be rewritten for Blackwell. (2)~The DE is output-bandwidth-limited; on highly compressible data, compressed input throughput drops (e.g., $1/C$ scaling), so DE use is most beneficial when decompressed output, not input bandwidth, is the bottleneck. (3)~PTX \texttt{tcgen05.mma} does not support FP64; FP64 DGEMM uses a different path (e.g., doubled FP64 units), so TMEM does not directly improve FP64 scientific kernels.

\textbf{Software Ecosystem:} CUDA 13.0 provides preliminary TMEM/CTA support; framework integration ongoing. FP6 hardware support exists but lacks software tooling. FP4/FP6 require per-layer precision selection—8.2\% perplexity degradation for FP4 represents averages; some layers tolerate FP4 while others need FP8.

\textbf{Performance guidelines.} We consolidate actionable recommendations: (a)~Use TMEM for accumulator staging and multi-stage tensor pipelines (e.g., fused attention $\mathbf{Q}\mathbf{K}^T$ then softmax then value) when working sets exceed L1; for single-shot small matrices, minimize TMEM round-trips. (b)~Scale DE usage with compression ratio: expect $\sim$170--220~GB/s decompressed output; compressed input rate scales inversely with ratio. (c)~Leverage warp-level \texttt{tcgen05.mma} and CTA-pair execution for GEMM and LLM inference; target 64$\times$64 tiles for TMEM. (d)~For LLM inference, FP8 balances throughput and quality; FP4 yields higher throughput with acceptable perplexity increase when validated per model.

\textbf{Deployment:} For LLM inference, B200 provides 1.8--3.9$\times$ advantages; FP4 practical for 70B models. Training improvements (1.85$\times$ ResNet-50, 1.55$\times$ GPT-1.3B) enable larger batches. HPC gains (1.92$\times$ FP64) competitive for scientific computing.

\section{Conclusion}
%This work presented a detailed experimental analysis of NVIDIA's Blackwell architecture (B200 chip) through carefully designed open-source microbenchmark suite. By comparing microarchitectural features of B200 against the Hopper (H200 chip) GPU, we provide insights into Blackwell's advancements in memory hierarchy, SM execution pipeline, and its 5th-gen Tensor Cores. Our analysis highlights the hardware's increased support for low-precision formats such as FP4 and FP6, revealing their practical implications for power and performance efficiency. The guidelines and observations presented in this study provide a microarchitectural understanding to assist developers in optimizing software to effectively use the hardware and thus enable more efficient deployment of AI and HPC workloads. 

NVIDIA's B200 GPU marks a major shift in GPU architectures. Our work presents the first detailed microbenchmark suite-based characterization of the NVIDIA Blackwell B200 GPU. Our work offers insights into its architectural innovations and performance behavior. We quantify the impact of TMEM on matrix-heavy workloads, evaluate the throughput and optimal usage of the hardware decompression engine, and analyze 5th-generation tensor core execution via the new \texttt{tcgen05} PTX instructions. Our study further assesses FP4 and FP6 precision trade-offs, benchmarks Blackwell across diverse workloads—including LLM inference, scientific kernels, and mixed-precision training—and distills actionable performance guidelines for developers targeting this next-generation architecture. Our experiments primarily exercised single-die behavior; dual-die (NV-HBI) interactions and cross-die latency remain as future work.

%Existing architectural simulators (Accel-Sim, GCoM) lack models for TMEM and Decompression Engine behavior, and vendor specifications provide only peak throughput numbers without latency distributions or resource contention analysis.

\section*{Acknowledgment}

This research used resources from the NVIDIA Brev Cloud Compute and the Google Cloud Provider. This material is based upon work supported by the U.S. DOE under Contract DE-FOA-0003177, S4PST: Next Generation Science Software Technologies Project.

% \newpage
% \clearpage

% % \section*{Artifact Description for IPDPS 2026 Reproducibility Initiative}
% \input{Appendix/ad}
% \section*{Artifact Evaluation (Optional)}
% \input{Appendix/ae}

\bibliographystyle{IEEEtran}
\bibliography{references}

% \appendix
% \input{Appendix/ipdps26_ad_ae_template}

% \input{Appendix/ipdps26_ad_ae_blackwell}

\end{document}